\documentclass[11pt]{article}
\usepackage{graphicx}

\begin{document}

\def\xslash#1{{\rlap{$#1$}/}}
\def \p {\partial}
\def \dd {\psi_{u\bar dg}}
\def \ddp {\psi_{u\bar dgg}}
\def \pq {\psi_{u\bar d\bar uu}}
\def \jpsi {J/\psi}
\def \psip {\psi^\prime}
\def \to {\rightarrow}
\def\bfsig{\mbox{\boldmath$\sigma$}}
\def\DT{\mbox{\boldmath$\Delta_T $}}
\def\xit{\mbox{\boldmath$\xi_\perp $}}
\def \jpsi {J/\psi}
\def\bfej{\mbox{\boldmath$\varepsilon$}}
\def \t {\tilde}
\def\epn {\varepsilon}
\def \up {\uparrow}
\def \dn {\downarrow}
\def \da {\dagger}
\def \pn3 {\phi_{u\bar d g}}

\def \p4n {\phi_{u\bar d gg}}

\def \bx {\bar x}
\def \by {\bar y}

\begin{center}
{\Large  On Unique Predictions for Single Spin Azimuthal  Asymmetry } \vskip
10mm
J. P. Ma and Q. Wang    \\
{\small {\it Institute of Theoretical Physics, Academia Sinica,
Beijing 100080, China }} \\
\end{center}

\vskip 1cm

\begin{abstract}
Theoretically there are two approaches to predict  single spin azimuthal asymmetries.
One is to take transverse momenta of partons into account
by using transverse momentum dependent parton distributions, while
another is to take asymmetries as a twist-3 effect. The nonperturbative
effects in these approaches are parameterized with different matrix elements
and predictions can be different. Recently, gauge invariant definitions
of transverse momentum dependent parton distributions were derived. With
these definitions
it can be shown that there are relations between nonperturbative matrix elements
in two approaches. These relations may enable us
to unify two approaches and to have unique predictions
for single spin azimuthal asymmetries.
In this letter we derive these relations by
using time-reversal symmetry and show that even with these relations
the single spin azimuthal asymmetry in Drell-Yan process is predicted
differently in different approaches.
\end{abstract}
\vskip 1cm

\eject\vfil
\par
Single spin azimuthal asymmetry provides a new tool to
study structure of hadrons because the asymmetry is sensitive
to correlations between quarks and gluons as partons inside
a hadron and to orbital angular momenta of these partons.
Experimentally,
such an asymmetry was observed  for inclusive production of pion
in polarized proton antiproton scattering with center-of-mass energy
$\sqrt{s}=20GeV$ by E704 collaboration~\cite{E704}.
The asymmetry is large for charged pion, while
for $\pi^0$ production it is consistent with zero when the transverse momentum
$k_T$ is smaller than 3GeV,
and it tends to a positive value when $k_T$ becomes large.
In semi inclusive deep-inelastic scattering(SIDIS) significant asymmetries were also
observed in production of pion and kaon by HERMES\cite{hermes}.
Asymmetries in polarized proton scattering are currently studied by STAR at RHIC.
Large spin effects are observed in their preliminary results after first run.
In SIDIS
measurements of asymmetries with a transversely polarized target were reported
by the SMC collaboration~\cite{BravarUT}.
Experiments with a transversely polarized target are
now under study by HERMES and COMPASS~\cite{HERMESPLAN,COMPASSPLAN}.
\par
Single spin azimuthal asymmetry is a T-odd effect and helicity-flip amplitudes
are involved. Perturbatively T-odd effects  can be generated at loop-level
in hard scattering of active partons of hadrons.
Because the quark-gluon coupling of QCD conserves
helicities in the massless limit, the T-odd effects are proportional
to quark masses which can be neglected. Therefore the observed T-odd effects
can not be explained by those T-odd effects arising from hard scattering
and are related to nonperturbative nature of hadrons.
Indeed, these T-odd effects can be generated from final- or initial interactions
between active partons involved in the hard scattering
and remnant partons in hadrons\cite{MSivers1,Col02, TMDJi}.
The effect of these interactions can be represented by gauge links in definitions
of parton distributions\cite{Col02, TMDJi}.
Theoretically there are two approaches to explain single spin azimuthal asymmetry
by taking
nonperturbative nature of hadrons into account.
One is to take transverse momenta $k_T$ of partons in a hadron into account
where one uses transverse momentum dependent parton distributions
to parameterize nonperturbative effects.
For a polarized
hadron as an initial state the effect is parameterized
by Sivers function\cite{Sivers}, while for a hadron observed in a final state
the T-odd effect related to this hadron
is parameterized by Collins function\cite{Collins93}.
For semi-inclusive deep inelastic scattering, both functions
can make contributions to the observed single spin azimuthal asymmetry.
Single spin azimuthal asymmetry has been studied in terms of these functions
\cite{Anselmino,Mulders,DeSanctis,Efremov,BQMa,Liang}.
These functions
have been also studied with models\cite{MCollins, MSivers1,MSivers2,Fyuan}.
Another approach, called Qiu-Sterman mechanism,
is that the T-odd effect is produced by taking twist-3 effect into account
and it is proportional to quark-gluon correlations inside a hadron\cite{Qiu}.
The fact that T-odd effects can be generated at twist-3 level
was also pointed out in \cite{EFTE}. This approach was used
to make predictions for various processes in \cite{Qiu,kako}.
It is interesting to note that at first look the physical reason for the effect
is different in different approaches. In the first approach the helicity
of a initial hadron is changed because of orbital angular momenta
of partons. This can be seen clearly in terms of light-cone wave functions\cite{JMY}.
In the second approach the helicity flip is caused by nonzero spin of the gluon which
is correlated with other partons. Predictions based on different approaches
are different. A question arises why there are two physical origins
for one effect?
\par
This question has been answered partly by recent studies of transverse momentum
dependent parton distributions\cite{TMD,Col02,TMDJi}, which are involved in the first
approach. It has been shown that gauge links in these distributions
play an important role to incorporate T-odd effects introduced by final state
interactions. In particular, additional gauge links should be included
in the definitions of these distributions\cite{TMDJi}. With these gauge links
it is possible to relate the second $k_T$ moment of Sivers function
to the twist-3 matrix element in the second approach\cite{Boer03}.
With relations between nonperturbative matrix elements in different
approaches it may be possible to unify two approaches and to
have unique predictions for single spin azimuthal asymmetries.
In this letter we will show that predictions based
on two approaches are still different, although such relations exist.
We will show this in detail with Drell-Yan process.
Before showing this we give another derivation of relations
between second $k_T$ moments of T-odd distributions and
twist-3 matrix elements by using time-reversal symmetry of QCD.
\par
We consider a proton moving in the $z$-direction with the momentum $P$
and the transverse spin ${\bf s_T}$. We use a light-cone coordinate
system and introduce two light-cone vectors: $n^\mu =(0,1,0,0)$,
$l^\mu =(1,0,0,0)$ and $n\cdot l =1$. Neglecting the proton mass  we have
$P^\mu =(P^+,0,0,0)$.
Taking transverse momenta of partons in quark-quark correlation in a proton
into account, there are two T-odd parton distribution functions appearing
in a Drell-Yan process which take effects of initial state interaction
into account. They can be defined
as\cite{Mulders97}
\begin{eqnarray}
f_{1T,DY}^\bot (x,k^2_T)
\varepsilon_{\perp\mu\nu} k^\mu_T s_T^\nu
  & =& \frac{1}{4}  \int \frac{d\xi^- d^2 \xi_T}{(2\pi)^3}
                e ^{-ik\cdot \xi}
               \{
                \langle P,{\bf s_T} \vert
                \bar\psi (\xi ) \gamma^+ V(\xi) \psi(0) \vert P,{\bf s_T} \rangle
 \nonumber\\
         &&          - ({\bf s_T} \to -{\bf s_T} ) \},
  \nonumber\\
h_{1,DY}^\bot (x,k^2_T) k_T^ i
   &=&  -\int \frac{d\xi^- d^2 \xi_T}{(2\pi)^3}
                e^{-ik\cdot \xi}
              \cdot
                \langle P \vert
                \bar\psi (\xi ) \sigma^{+i} V(\xi) \psi(0) \vert P \rangle,
\end{eqnarray}
with $\xi^\mu =(0,\xi^-, {\bf \xi_T})$ and
$\varepsilon_{\perp\mu\nu}=\varepsilon_{\rho\sigma\mu\nu} n^\rho l^\sigma$.
The momentum $k$ is $k^\mu =(xP^+,0,{\bf k_T})$.
The matrix element in the last line is spin averaged.
The function $f_{1T,DY}^\bot (x,k^2_T)$ is the Sivers function
for Drell-Yan processes.
$V(\xi )$
is a product of gauge links to make the matrix element gauge invariant,
it takes the effect of initial state interaction in Drell-Yan process
into account.
If one can take $V(\xi)$ as a unit matrix, then one can show with time-reversal symmetry
that both correlation functions are zero. It is
important to  note that $V(\xi)$ is not a unit matrix even
in the light-cone gauge $n\cdot G=0$, additional gauge links must be introduced
to make the definitions gauge invariant\cite{TMDJi}.
We will take the light-cone gauge. In this gauge $V(\xi)$ reads:
\begin{equation}
  V(\xi) = V_{-\infty} ({\bf \xi_T} ) = P {\rm exp}
      \left ( ig \int_0^{{\bf \xi_T}} d {\bf \xi_T}
      \cdot {\bf G_T } (0,\xi^-=-\infty, {{\bf \xi_T}})
      \right ).
\end{equation}
This gauge link takes effects of initial state interaction into account and
it can be derived in a similar way as in SIDIS\cite{TMDJi}. The difference
is that the gauge link is at $\xi^-=-\infty$ because it is for initial state
interaction.
\par
Under parity- and time-reversal transformation, we obtain for the matrix
element:
 \begin{equation}
 \label{eqn1}
 \langle P,{\bf s_T } \vert
               \bar\psi (\xi )
                \gamma^+ V_{-\infty}({\bf \xi_T}) \psi(0) \vert P,{\bf s_T } \rangle
       =\langle P,-{\bf s_T } \vert
                \bar\psi (\xi )
                \gamma^+ V_{\infty}({\bf \xi_T}) \psi(0) \vert P,-{\bf s_T } \rangle,
\end{equation}
with
\begin{equation}
 V_{\infty} ({{\bf \xi_T} }) = P {\rm exp}
      \left ( ig \int_0^{{\bf \xi_T} } d{{\bf \xi_T} }\cdot
      {\bf G_T } (0,\xi^-=\infty, {\bf \xi_T} )
      \right ).
\end{equation}
Similarly one can define two T-odd parton distribution functions appearing
in deep inelastic processes which take effects of final state interaction
into account. The two functions $f_{1T,DIS}^\bot (x,k^2_T)$ and
$h_{1,DIS} ^\bot (x,k^2_T)$ are defined similarly as in Eq.(1), but with
the gauge link $V(\xi)$ is replaced with $V_{\infty} ({{\bf \xi_T} })$. These two
functions are related to those in Drell-Yan processes with time-reversal symmetry.
With Eq.(~\ref{eqn1}) we can write:
\begin{eqnarray}
\label{eqn2}
f_{1T,DY}^\bot (x,k^2_T)
\varepsilon_{\perp\mu\nu} k^\mu_T s_T^\nu
  & =& \frac{1}{8}  \int \frac{d\xi^- d^2 \xi_T}{(2\pi)^3}
                e ^{-ik\cdot \xi}
               \{
                \langle P,{\bf s_T} \vert
                \bar\psi (\xi ) \gamma^+
  \left [ V_{-\infty} ({\bf \xi_T} ) - V_{\infty}  ({\bf \xi_T} )\right ]
   \psi(0) \vert P,{\bf s_T} \rangle
 \nonumber\\
         &&          - ({\bf s_T} \to -{\bf s_T} ) \},
  \nonumber\\
h_{1,DY} ^\bot (x,k^2_T) k_T^ i
   &=&  -\frac{1}{2}  \int \frac{d\xi^- d^2 \xi_T}{(2\pi)^3}
                e ^{-ik\cdot \xi }
                \langle P \vert
                \bar\psi (\xi ) \sigma^{+i}
                \left [ V_{-\infty} ({\bf \xi_T} ) - V_{\infty}  ({\bf \xi_T} )\right ]
                 \psi(0) \vert P \rangle.
\end{eqnarray}
It is expected that the function $f_{1T,DY}^\bot$ and $h_{1,DY}^\bot$
decrease rapidly with increasing $k_T$.
Then T-odd effects related to them can be estimated
at leading order by the second moment of $k_T$ of the left hand side
in Eq.(5):
\begin{eqnarray}
K_f^\alpha (x) & =&  \int d^2 k_T  k_T ^\alpha
f_{1T,DY}^\bot (x,k^2_T)
\varepsilon_{\perp\mu\nu} k^\mu_T s_T^\nu
   =-\frac{1}{2} \varepsilon^{\alpha\sigma}s_{T\sigma}
    \int d^2k_T \vert {\bf k_T }\vert ^2 f_{1T,DY}^\bot (x,k^2_T),
\nonumber\\
K_h^{\mu\nu}(x)  & =&  \int d^2 k_T  k_T ^\mu
h_{1,DY}^\bot (x,k^2_T) k_T^ \nu
    =\frac{1}{2} (n^\mu l^\nu+n^\nu l^\mu -g^{\mu\nu} )
    \int d^2 k_T\vert {\bf k_T }\vert ^2 h_{1,DY}^\bot (x,k^2_T).
\end{eqnarray}
With Eq.(~\ref{eqn2}) these moments can be expressed in term of matrix elements.
Taking $K_f^\alpha$ as an example, we have
\begin{eqnarray}
K_f^\alpha(x) & =&-\frac{1}{8}  \int \frac{d\xi^- }{(2\pi)}
                e ^{-i xP^+ \xi^- }
               i \frac{\partial }{\partial \xi_{T\alpha}} \{
                \langle P,{\bf s_T} \vert
                \bar\psi (\xi ) \gamma^+
  \left [ V_{-\infty} ({\bf \xi_T} ) - V_{\infty}  ({\bf \xi_T} )\right ]
   \psi(0) \vert P,{\bf s_T} \rangle
    \nonumber\\
         &&          - ({\bf s_T} \to -{\bf s_T} ) \}\vert_{{\bf \xi_T} =0}.
\end{eqnarray}
Carrying out the derivatives we have:
\begin{eqnarray}
 K_f^\alpha (x)  &=& \frac{1}{8}  \int \frac{d\xi^- }{(2\pi)}
                e^{-ix P^+\xi^- }
 \cdot \{g \langle P,{\bf s_T } \vert
                \bar\psi (\xi^-n) \gamma^+ [
                G_T^\alpha  (0,\infty,0,0) -
   G_T ^\alpha (0,-\infty,0,0) ] \psi(0) \vert P,{\bf s_T } \rangle
\nonumber \\
   && \ \ \ - ({\bf s_T } \to -{\bf s_T }) \}.
\end{eqnarray}
\par
Now one can show that $K^\alpha$ is related to
the twist-3 quark gluon correlation $T_F(x,x)$ introduced in \cite{Qiu}.
The correlation function
is defined as:
\begin{eqnarray}
T_F (x_1,x_2) \epsilon^{\mu\nu\sigma\rho} n_\nu l_\sigma s_{T\rho}
   &=&  -\frac{g}{2}\int \frac{dy_1 dy_2}{4\pi}
   e^{ -iy_2 (x_2-x_1) P^+ -i y_1 x_1 P^+ }
   \nonumber\\
    && \cdot  \{ \langle P,{\bf s_T } \vert
           \bar\psi (y_1n ) \gamma^+ G^{+\mu}(y_2n) \psi(0) \vert P,{\bf s_T } \rangle
  - ({\bf s_T } \to -{\bf s_T }) \}.
\end{eqnarray}
where we include the coupling constant $g$ into the definition. It is straightforward
to obtain:
\begin{equation}
\label{eqn3}
  T_F(x,x) =\int d^2 k_T \vert {\bf k_T }\vert ^2 f_{1T,DY}^\bot (x,k^2_T).
\end{equation}
Similarly we have:
\begin{equation}
  T_H(x,x) = \int d^2 k_T \vert {\bf k_T }\vert ^2 h_{1,DY}^\bot (x,k^2_T),
\end{equation}
where $T_H$ is defined with a twist-3 operator:
\begin{equation}
T_H (x_1,x_2)
   = g \int \frac{dy_1 dy_2}{4\pi}
    e^{ -iy_2 (x_2-x_1) P^+ -i y_1 x_1 P^+ }
     \langle P \vert
           \bar\psi (y_1n ) \sigma^{+\mu}
           G^+_{\ \mu} (y_2n) \psi(0) \vert P \rangle.
\end{equation}
The relations in Eq.(10,11) clearly show that the effect of orbital
angular momenta of quarks is closely related to that of quark-gluon correlations
because of gauge invariance. These relations also show
that the nonperturbative effects in the two approaches
for single spin azimuthal asymmetries are the same. However, it should
be noted that perturbative coefficients
in these two approaches are calculated in different ways.
In the first approach one uses $k_T$-factorization, while
collinear expansion is used in the second approach. If the
perturbative coefficients in the two approaches are related in
a consistent way so that the predicted single spin azimuthal asymmetries
are same,
then we may have an unique prediction for single spin azimuthal asymmetries
and the question asked before is fully answered. It is difficult
to establish a general relation between the perturbative coefficients,
since they are differently calculated in different ways and are different
in different processes.
But we can show that the single spin azimuthal asymmetry in Drell-Yan
process is {\it differently} predicted between the two approaches.
\par
Now we calculate single spin asymmetry in Drell-Yan process:
\begin{equation}
  A(P_A,s_T) + B(P_B) \to l^-(P_1) +l^+(P_2) +X,
\end{equation}
where the proton A is
transversely polarized with the spin vector ${\bf s_T}$
and moves in the $+z$-direction. The $x$-direction is chosen
as the direction of ${\bf s_T}$. $S=(P_A+P_B)^2$. The hadron $B$ is unpolarized
and moves in $-z$-direction. We will calculate the single spin azimuthal asymmetry
at leading orders, where the lepton pair has a small transverse momentum.
We assume that the solid
angle $\Omega(\theta,\phi)$ of the produced lepton in the center-of-mass frame
of the produced lepton pair and the invariant mass $Q^2$
of the lepton pair is observed.
The single spin asymmetry is defined as:
\begin{equation}
A_N =  \left ( \frac{d\sigma({\bf S_T})}{dQ^2d\Omega}
  -\frac{d\sigma({-\bf S_T })}{dQ^2d\Omega} \right ) \Big /
   \left ( \frac{d\sigma({\bf S_T})}{dQ^2d\Omega}
  +\frac{d\sigma({-\bf S_T})}{dQ^2d\Omega} \right ).
\end{equation}
The asymmetry is calculated in \cite{DY} with Qiu-Sterman mechanism.
The result reads:
\begin{equation}
A_N = -\frac{1}{Q}\cdot \frac{\sin 2\theta \sin \phi}{1+\cos^2\theta}
\cdot \left ( \frac {\sum_q e_q^2 \int dx_A dx_B \delta(Q^2-x_A x_B S)
         T_{F,q/A}(x_A,x_A) f_{\bar q/B}
              ( X_B) }
             {\sum_q e_q^2 \int dx_A dx_B f_{q/A}(x_A) f_{\bar q/B}
              ( x_B) }  + \cdots \right )
\end{equation}
where $T_{F,q/P}(x,x)$ is defined in Eq.(9). The subscriber $q/A$
denotes the distribution of $q$ in hadron $A$. It should be noted
that the summation $\sum_q$ and also in the below
is over all quark and antiquark flavors, i.e., $q$ can be an antiquark in
the summation.
We only keep the term
with $T_{F,q/A}$. The $\cdots$ represents another term proportional to $T_{H,\bar q/B}$
which is irrelevant in this letter.
\par
In order to make comparison of two approaches we study
the single spin asymmetries with $k_T$ dependent distributions.
At tree level
the partonic process is just $q\bar q\to l^+ l^-$.
The cross section can be written:
\begin{equation}
\sigma = e^2 \frac{1}{2S } \int \frac{d^3 P_1}{(2 \pi)^3 2 P^0_1}
\frac{d^3 P_2}{(2 \pi)^3 2 P^0_2} L_{\mu \nu} W^{\mu \nu}\cdot\frac{1}{Q^4}
\end{equation}
where the leptonic tensor $L_{\mu \nu}$ and hadronic tensor $W^{\mu \nu}$ are given by:
\begin{eqnarray}
L^{\mu\nu} &=& 4(P_1^\mu P_2^\nu+P_1^\nu P_2^\mu-P_1\cdot P_2 g^{\mu\nu}),
\nonumber\\
W^{\mu \nu} &=& \sum_q e_q^2 \int \frac{d^4 k_A}{(2 \pi)^4} \frac{d^4 k_B}{(2 \pi)^4}
(2 \pi)^4 \delta^4 (Q-k_A -k_B)
\int d^4 \xi_1 d^4 \xi_2 e^{i k_A \cdot \xi_1 + i k_B \cdot \xi_2} \cdot (\gamma^\nu)_{jk} \cdot (\gamma^\mu)_{li}
\nonumber \\
  && \cdot \left [ \langle P_A,s_T \vert \bar q_j (0) {q_{i}} (\xi_1) \vert P_A,s_T \rangle
  \langle P_B \vert {q_k} (0) \bar{q_l} (\xi_2) \vert P_B \rangle
  + \cdots \right ],
\end{eqnarray}
where the $\cdots$ denotes power-suppressed terms.
The quark $q$ and $\bar q$
carries the momentum
\begin{eqnarray}
k_A= x_A P_A + k_{AT} \nonumber \\
k_B=x_B P_B + k_{BT}
\end{eqnarray}
respectively.
The quark density matrix with $k_T$ dependence
\begin{equation}
\Phi_{ij} (x,k_T;P,S) = \frac{1}{2}
\int \frac{d\xi^- d^2\xi_T}{(2 \pi)^3} e^{i k\cdot \xi} \langle P,S \vert \bar \psi_j (0) \psi_i (\xi) \vert P,S \rangle |_{\xi^+ =0}
\end{equation}
can be parameterized as\cite{Mulders97}
\begin{eqnarray}
\label{eqn5}
\Phi(x,k_T;P,S)&=& \frac{1}{4} \{ f_1 \xslash l  +
 f_{1T,DY}^\bot \epsilon_{\mu \nu \rho \sigma} \gamma^\mu l ^\nu k_T^\rho s_T^\sigma
 + g_{1s} \gamma_5 \xslash l  + h_{1T} i \sigma_{\mu \nu} \gamma_5 l ^\mu s_T^\nu
 \nonumber  \\
  && + h_{1s}^\bot i \sigma_{\mu \nu} \gamma_5 l ^\mu k_T^\nu +
  h_{1,DY}^\bot \sigma_{\mu \nu} k_T^\mu l ^\nu \}
\end{eqnarray}
where the Sivers function is $f_{1T,DY}^\bot$.
The function $f_{1T,DY}^\bot$ and $h_{1,DY}^\bot$ is defined in Eq.(1).
We changed the notation of \cite{Mulders97}
slightly by replacing $1/M$ with 1.
With this parameterization we have
\begin{eqnarray}
W^{\mu \nu}&=& \frac{1}{3} \sum_q e_q^2 \int dk_A^+ d^2k_{AT}
    dk_B^- d^2 k_{BT} (2\pi)^4 \delta^4(Q-k_A-k_B)
    ({\bf k_{AT}} \times {\bf s_T}) \cdot \hat z \nonumber \\
    &\ &\cdot f_{1T,q/A}^\bot (k_A) f_{1,\bar q/B}(k_B)
    [g^{\mu \nu} - l^\nu n^\mu - l^\mu n^\nu] + \cdots,
\end{eqnarray}
where we keep only terms with $f_{1T,DY}^\bot$. $\hat z$ denotes
the direction of the $z$-axis.
It should be noted that the total momentum $Q$ of the lepton pair
has nonzero transverse components in general. It depends
on transverse momenta of incoming partons.
It is now straightforward to calculate the asymmetry defined in Eq.(14).
Since the asymmetry is defined as a distribution of variables
in the center-mass frame of the lepton pair, we need to specify
the frame. We assume that the center-mass frame is obtained
from laboratory frame by a Lorentz boost only. This is conveniently used
in experiment. In the center-mass
frame the lepton $l^+$ and $l^-$ has the momentum $k_2$ and $k_1$
respectively. The momentum $k_1$ and $k_2$ read:
\begin{eqnarray}
k_1^\mu &=&\frac{\sqrt{Q^2}}{2} (1, \sin\theta\sin\phi,\sin\theta\cos\phi,\cos\phi),
\nonumber\\
k_2^\mu &=&\frac{\sqrt{Q^2}}{2} (1, -\sin\theta\sin\phi,-\sin\theta\cos\phi,-\cos\phi)
\end{eqnarray}
The momentum $P_i$$(i=1,2)$ in the laboratory frame is related to $k_i$$(i=1,2)$
by the boost:
\begin{eqnarray}
P^0_i &=& \frac{Q^0}{\sqrt{Q^2}} \left ( k_i^0
+\frac{{\bf Q}\cdot {\bf k}_i}{Q^0}\right ),
\nonumber\\
{\bf P}_i &=& {\bf k}_i +\frac{k^0_i}{\sqrt{Q^2}} {\bf Q}
                 +\left ( \frac{Q^0}{\sqrt{Q^2}}-1\right )
                \frac {{\bf Q}\cdot {\bf k}_i}{{\bf Q}\cdot {\bf Q}}
                  {\bf Q}.
\end{eqnarray}
The phase space integration is invariant under the boost.
Using Eq.(23) one can express $L^{\mu\nu}$ in term of $k_1$,$k_2$ and $Q$.
For transverse momentum independent parton distributions
one expects in general that they decrease rapidly with increasing
transverse momenta. Hence an expansion of the perturbative part
in transverse momenta is
an good approximation.
Keeping the first non-zero order in the expansion
of $k_{AT}$ and $k_{BT}$,
we obtain the asymmetry:
\begin{eqnarray}
  A_N &=& \frac{1}{Q}\cdot\frac{ \sin 2 \theta \sin \phi}{1+\cos^2\theta}
   \sum_q e_q^2  \int dx_A d x_B
\delta (Q^2-x_A x_B S)
  \frac{x_B - x_A }{2(\sqrt{x_A } + \sqrt{x_B})^2}
\nonumber\\
&& \cdot  f_{\bar q/B} (x_B)\cdot
\int d^2 k_T \vert{\bf k_T}\vert^2 f_{1T,DY,q/A}^\bot (x_A,k^2_T )
/{\sum_q e_q^2 \int dx_A dx_B f_{q/A}(x_A) f_{\bar q/B}( x_B) }
+\cdots.
\nonumber\\
&=& -\frac{1}{Q}\cdot\frac{ \sin 2 \theta \sin \phi}{1+\cos^2\theta}
   \sum_q e_q^2  \int dx_A d x_B
\delta (Q^2-x_A x_B S)
  \frac{x_A - x_B }{2(\sqrt{x_A } + \sqrt{x_B})^2}
\nonumber\\
&& \cdot  f_{\bar q/B} (x_B)\cdot
T_{F,q/P}(x_A,x_A)
/{\sum_q e_q^2 \int dx_A dx_B f_{q/A}(x_A) f_{\bar q/B}( x_B) }
+\cdots.
\end{eqnarray}
In the above equation we have assumed that the initial hadrons
are in a center-mass frame, i.e., $P_A^0=P_B^0$.
In the last step we have used the relation in Eq.(10).
Again, the summation $\sum_q$
is over all quark and antiquark flavors.
It is clearly that the asymmetry here is different
than that in Eq.(15) because of the factor
${(x_A - x_B) }/{2(\sqrt{x_A } + \sqrt{x_B})^2}$. If the factor was $1$, then
the asymmetry would be the same. Hence, the asymmetry
obtained by two approaches will have the same angular distribution
but the normalization is different.
Since the factor can be positive or
negative, the asymmetry from the two approaches can even have different signs.
\par
It should be noted that the hadronic tensor calculated
with the parameterization in Eq.(20) is not invariant under electromagnetic
gauge transformation. This can be seen by evaluating $Q_\mu W^{\mu\nu}$
with $W^{\mu\nu}$ given in Eq.(21).
The reason is that the partons involved in the hard scattering have nonzero transverse
momenta and $\gamma\cdot l=\gamma^-$ is contracted with the hard part of $q\bar q\to l^+l^-$
according to the first two terms in Eq.(20). With nonzero transverse
momenta the contraction with $\gamma^-$ does not make the $q\bar q$ pair
on-shell. Hence the $U_{em}(1)$ gauge invariance is not preserved.
It is also indicated\cite{Metz} that the decomposition in Eq.(20) may need to be
reexamined because the density matrix element also depends on the
vector $n^\mu$ due to gauge links.
This dependence is neglected in Eq.(20). There are possibly many ways
to restore the gauge invariance. In this letter we simply make the initial
parton on-shell by
replacing  $\gamma^-$ with $\gamma\cdot k/n\cdot k$
for the first two terms in Eq.(20), i.e.,
\begin{equation}
\Phi(x,k_T;P,s_T) = \frac{1}{4} \{ f_1(x,k_T^2) \frac{\gamma \cdot k}{k \cdot n}
+ f_{1T,DY}^\bot (x,k_T^2) \frac{\gamma \cdot k}{k \cdot n}
\varepsilon_{\perp\rho \sigma} k_T^\rho s_T^\sigma + \cdots.
\end{equation}
The hadronic tensor obtained with Eq.(25) is $U_{em}(1)$-gauge invariant.
It is:
\begin{eqnarray}
W^{\mu \nu}&=& \frac{1}{3} \sum_q e_q^2 \int dk_A^+ d^2k_{AT} dk_B^- d^2k_{BT}
    (2\pi)^4 \delta^4(Q-k_A-k_B) ({\bf k_{AT}} \times {\bf s_T})\cdot \hat z
\nonumber \\
        &\ &\frac{1}{(k_A \cdot n)(k_B \cdot l)} \big [{g^{\mu \nu} k_A \cdot k_B
    - k_A^\mu k_B^\nu - k_A^\nu k_B^\mu }\big ]  f_{1T,DY,q/A}^\bot(k_A) f_{1,\bar q/B}(k_B).
\end{eqnarray}
It is straightforward to show
$Q_\mu W^{\mu \nu} \sim k_A^2 k_B^\nu + k_B^2 k_A^\nu \sim k^2_T$.
Hence the gauge invariance is preserved up to order $k_T^2$. The asymmetry
calculated with this tensor will be gauge invariant. The result of $A_N$
can be obtained
from Eq.(24) by replacing the factor
${(x_B - x_A) }/{2(\sqrt{x_A } + \sqrt{x_B})^2}$ with
${\sqrt{x_B}  }/{(\sqrt{x_A } + \sqrt{x_B})}$. Therefore,
even after we make the hadronic tensor gauge invariant, the obtained
asymmetry $A_N$ is still different than that in Eq.(15) from the second
approach. It is interesting to see how the same asymmetry in Eq.(15)
can be obtained by starting from Eq.(26).  If we replace
the tensor $[g^{\mu\nu} k_A \cdot k_B
    - k_A^\mu k_B^\nu - k_A^\nu k_B^\mu ]$
with $[g^{\mu \nu} k_A \cdot k^\prime_B- k_A^\mu k_B^{\prime\nu}
- k_A^\nu k_B^{\prime\mu} ]$
where $k_B^{\prime\mu} =(0,k_B^-, -{\bf k_{AT}})$ and neglect
the dependence of lepton momenta on transverse momenta of incoming
partons, we indeed obtain the asymmetry in Eq.(15)
with the same normalization, but with an extra negative sign.
However, the transverse momentum of the lepton pair
can not be neglected and we can not do the replacement.
\par
To summarize: There are two different approaches
for single spin azimuthal asymmetries.
Using time-reversal symmetry, we give in this letter an detailed
derivation of the relations
between $k_T$ dependent T-odd distributions and twist-3 quark-gluon
correlators, which are used
in different approaches, respectively.
These relations show that the physical origin
in the two different approaches
for single spin azimuthal asymmetries is the same because
of the gauge invariance.
With these relations it may be expected to unify these two approaches
and to delivery an unique prediction for single spin azimuthal asymmetries.
We have studied in detail the single spin azimuthal asymmetry
in Drell-Yan process with Sivers function and found that predictions
from different approaches are different even with these relations.
The $k_T$ factorization
used for single spin azimuthal asymmetries does not
respect the $U_{em}(1)$-gauge invariance. This problem
may be solved by changing the projection of the perturbative part
slightly. But even after this changing the predicted
asymmetry is still different. Our study shows clearly
that different approaches give different predictions
for the same physical effect in Drell-Yan process and one can
expect that the same situation  will also appear in other processes.
Therefore, at present we
have not an unique prediction for single spin azimuthal asymmetries
in Drell-Yan process at least
and  this problem needs to be studied further.
\par\vskip20pt
{\bf Acknowledgements}
\par
We would like to thank Prof. X.D. Ji and Prof. J.W. Qiu
for useful discussions.
This work is supported by National Nature
Science Foundation of P. R. China.

\par\vfil\eject

\end{document}